\documentclass[12pt]{iopart}

\usepackage{graphicx}
\usepackage{url}
\usepackage{wasysym}

\newcommand{\ve}[1]{\mbox{\boldmath$#1$}}

\def\source{{\rm 0}}
\def\obs{{\rm 1}}

\let\oldbibitem\bibitem
\renewcommand\bibitem[2][]{\oldbibitem{#2}}

\begin{document}

\title{A generalized lens equation for light deflection in weak gravitational fields}

\author{Sven Zschocke}

\address{Lohrmann Observatory, Dresden Technical University,
Mommsenstr. 13, 01062 Dresden, Germany}

\ead{Sven.Zschocke@tu-dresden.de}

\begin{abstract}

A generalized lens equation for weak gravitational fields in 
Schwarzschild metric and valid for finite distances of source and observer
from the light deflecting body is suggested. 
The magnitude of neglected terms in the generalized lens equation is estimated to be smaller than or equal 
to $\displaystyle \frac{15\,\pi}{4}\,\frac{m^2}{{d^{\;\prime}}^2}$, where 
$m$ is the Schwarzschild radius of massive body and $d^{\;\prime}$ is Chandrasekhar's impact parameter. 
The main applications of this generalized lens equation are extreme astrometrical configurations, 
where {\it Standard post-Newtonian approach} as well as {\it Classical lens equation} cannot be applied. 
It is shown that in the appropriate limits the proposed lens equation yields the known post-Newtonian terms, 
'enhanced' post-post-Newtonian terms and the Classical lens equation, thus provides a link between these both 
essential approaches for determining the light deflection. 

\end{abstract}

\pacs{95.10.Jk, 95.10.Ce, 95.30.Sf, 04.25.Nx, 04.80.Cc}

\section{Introduction}

Todays astrometry necessitates theoretical predictions of light-deflection by massive bodies on microarcsecond
($\mu{\rm as}$) level, e.g. astrometric missions SIM (NASA) or GAIA (ESA). In principle, an
astrometric precision on microarcsecond level can be achieved by numerical integration of geodesic equation of
light-propagation. On the other side, modern astrometric missions like GAIA determine the positions and proper motions of
approximately one billion objects, each of which is observed about one hundred times.
The data reduction of such huge amount of observations implies the need of analytical solutions, because the numerical
investigation of geodesic equation is by far too
time-consuming.

The metric of a massive body can be expanded in terms of multipoles, i.e. monopole term, quadrupole
term and higher multipoles \cite{Thorne,Blanchet}. 
Usually, the largest contributions of light-deflection originates from the spherically
symmetric part (Schwarzschild) of the massive body under consideration. The exact analytical solution of light-propagation
in Schwarzschild metric \cite{Chandrasekhar1983} inherits elliptic integrals, but their evaluation becomes comparable with
the time effort needed for a numerical integration of geodesic equation. Thus, approximative analytical solutions
valid on microarcsecond level of accuracy are indispensible for a highly time-efficient data reduction.

In the same way, exact lens equations of light-deflection have been obtained in \cite{exact_lens1,exact_lens2,exact_lens3}.
However, such exact relations are also given in terms of elliptic integrals. Therefore, approximations of these exact 
solutions are also needed for a time-efficient data reduction. 
An excellent overview of such approximative lens equations has recently been presented in \cite{exact_lens4}.

Basically, two essential approximative approaches are known in order to determine the light-deflection 
in weak gravitational fields:

The first one is the standard parameterized post-Newtonian approach (PPN) 
\cite{Misner_Thorne_Wheeler,Brumberg2} which is of the order ${\cal O} \left(m\right)$. During the last
decades, it has been the common understanding that the higher order terms ${\cal O} \left(m^2\right)$ are negligible even
on microarcsecond level, except for observations in the vicinity of the Sun. However, recent investigations
\cite{Teyssandier_LePoncinLafitte,Ashby_Bertotti,Article_Klioner_Zschocke,Teyssandier} have revealed that the
post-post-Newtonian approximation \cite{Brumberg1,Brumberg2}, which is of the order ${\cal O} \left(m^2\right)$, is needed
for such high accuracy. Both approximations are applicable for $d \gg m$, where $d$ being the impact parameter of the
unperturbed light ray.

The second one is the standard weak-field approximative lens equation, which is usually called the classical lens equation,
see Eq.~(67) in \cite{exact_lens3} of Eq.~(24) in \cite{exact_lens4}.
One decisive advantage of classical lens equation is its validity for arbitrarily small values of impact parameter $d$.
The classical lens equation is valid for astrometrical configurations where source and observer are far ernough from
the lens, especially in case of $a \gg d$ and $b \gg d$, where $a = \ve{k}\cdot\ve{x}_{\obs}$ and
$b = - \ve{k}\cdot\ve{x}_{\source}$, where $\ve{x}_{\source}$ and $\ve{x}_{\obs}$ are the three-vectors from the
center of the massive body to the source and observer, respectively, and $\ve{k}$ is the unit vector from the
source to the observer. However, the classical lens equation is not applicable for determine 
the light-deflection of moons of giant planets in the solar system, because astrometrical configurations 
with $b = 0$ are possible. 

Moreover, there are astrometric configurations where neither the standard post-Newtonian approach nor the classical lens 
equation are applicable, for instance binary systems. In order to investigate the light-deflection in such systems
a link between these both approaches is needed. Such a link can be provided by a generalized lens equation which,
in the appropriate limits, coincides with standard post-Newtonian approach and classical lens equation.
Accordingly, the aim of our investigation is an analytical expression for the generalized lens equation
having a form very similar to the classical lens equation. We formulate the following conditions under which our
generalized lens equation should be applicable:

\begin{enumerate}
\item[] $1.$ valid for $d = 0\;,\; a = x_{\obs} \gg m\;,\;b = x_{\source} \gg m$,
\item[] $2.$ valid for $a = 0\;,\;d \gg m\;,\;b \neq 0$,
\item[] $3.$ valid for $b = 0\;,\;d \gg m\;,\;a \neq 0$.
\end{enumerate}

These conditions imply that the light-path is always far enough from the lens, thus inherit weak gravitational fields,
i.e. small light deflection angles.
In order to control the numerical accuracy, the generalized lens equation is compared with the numerical solution of
exact geodesic equation in the Schwarzschild metric (throughout the paper, we work in harmonic gauge):
\begin{eqnarray}
g_{00} &=& - \frac{1-a}{1+a}\;,\quad \quad g_{i0} = 0\,,
\nonumber\\
g_{ij} &=& \left(1+a\right)^2\delta_{ij} + \frac{a^2}{x^2}\,\frac{1+a}{1-a}\,x^i\,x^j\,.
\label{exact_Schwarzschild_5}
\end{eqnarray}

\noindent
Here, $\displaystyle a=\frac{m}{x}$ and $\displaystyle m=\frac{G\,M}{c^2}$ is the Schwarzschild radius
and $M$ is the mass of the light-deflecting body, $G$ is Newtonian constant of gravitation and $c$ is the speed of light.
Latin indices take values $1,2,3$, and the Euclidean metric $\delta_{ij}=1 (0)$ for $i=j$ ($i\neq j$).
The absolute value of a three-vector is denoted by $x=\left|\ve{x}\right| = \sqrt{x_1^2 + x_2^2 + x_3^2}$.
The exact geodesic equation in Schwarzschild metric reads, cf. \cite{Article_Klioner_Zschocke}
\begin{eqnarray}
\fl 
\ddot{\ve{x}} =
\frac{a}{x^2} \left[ - c^2 \frac{1 - a}{(1 + a)^3} - \dot{\ve{x}} \cdot \dot{\ve{x}}
+ a \frac{2 - a}{1 - a^2} \left( \frac{ {\ve{x}} \cdot \dot{\ve{x}}}{x} \right)^2
\right] \ve{x} +  2 \frac{a}{x^2} \;\frac{2 - a}{1 - a^2}
( \ve{x} \cdot \dot{\ve{x}}) \, \dot{\ve{x}} \,,
\label{exact_Schwarzschild_10}
\end{eqnarray}

\noindent
where a dot denotes time derivative in respect to the coordinate time $t$, and $\ve{x}$ is the three-vector
pointing from the center of mass of the massive body to the photon trajectory at time moment $t$.
The scalar product of two three-vectors with respect to Euclidean metric $\delta_{ij}$ is
$\ve{a}\cdot\ve{b}=\delta_{i j}\, a^i b^j$.
The numerical solution of this equation will be used in order to determine the accuracy of approximative solutions.
We abbreviate the angle between two three-vectors $\ve{a}$ and $\ve{b}$ by $\delta(\ve{a},\ve{b})$,
which can be computed by means of $\displaystyle \delta(\ve{a},\ve{b}) = \arccos \frac{\ve{a}\cdot\ve{b}}{a\,b}$.

The paper is organized as follows: In Section \ref{Standard_post_Newtonian} 
the post-Newtonian approach is presented. 
In Section \ref{post_post_Newtonian} the steps of post-post-Newtonian approach
relevant for this investigation are shown, and I will briefly summarize the 
main results of our article \cite{Article_Klioner_Zschocke}.
The generalized lens equation is obtained in Section \ref{generalized_lens} and
discussed in Section \ref{discussion}. A summary is given in Section \ref{summary}.

\section{Post-Newtonian approximation\label{Standard_post_Newtonian}}

Let us consider the trajectory of a light-signal in post-Newtonian Schwarzschild metric: 
\begin{eqnarray}
g_{00} &=& - 1 + 2\,a + {\cal O} \left(c^{-4}\right)\;, \quad g_{i0}=0\,,
\nonumber\\
g_{ij} &=& \delta_{ij} + 2\,\gamma\,a\,\delta_{ij} + {\cal O} \left(c^{-4}\right). 
\label{pN_mestric_5}
\end{eqnarray}

\begin{figure}[!h]
\begin{center}
\includegraphics[scale=0.8]{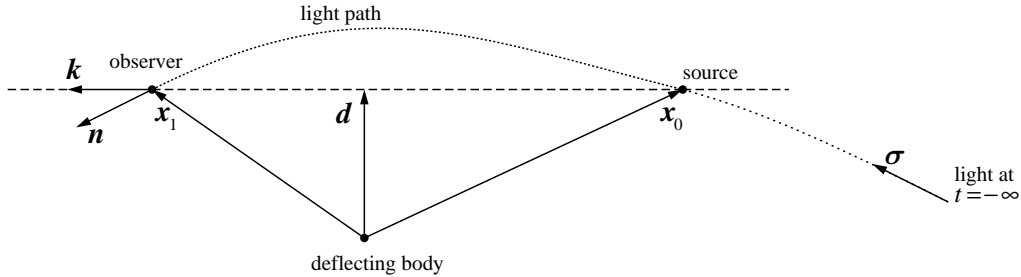}
\caption{A geometrical representation of the boundary problem under consideration for a light-propagation from the 
source to the observer.} 
\label{Fig: LensC}
\end{center}
\end{figure}

\noindent
Here, $\gamma$ is the parameter of the Parametrized Post-Newtonian (PPN) formalism, which characterizes 
possible deviation of the physical reality from general relativity theory where $\gamma=1$.
The light-ray is being emitted at a position $\ve{x}_{\source}$ at time moment $t_0$ and received at position 
$\ve{x}_{\obs}$ at a time moment $t_1$, see Figure~\ref{Fig: LensC}.

\noindent
Light propagation is governed by the geodesic equation, in post-Newtonian order given by 
\begin{eqnarray}
\ddot{\ve{x}} &=&
-(1+\gamma)\,c^2\,\frac{a\,\ve{x}}{x^2}
+2\,(1+\gamma)\,\frac{a\,\dot{\ve{x}}\,(\dot{\ve{x}}\cdot\ve{x})}{x^2} + {\cal O} (c^{-2})\,.
\label{geodesic-post-Newtonian}
\end{eqnarray}

\noindent
The unit tangent vector at the point of observation is 
$\displaystyle \ve{n} = \frac{\dot{\ve{x}}(t_1)}{\left|\dot{\ve{x}}(t_1)\right|}$, and the unit 
tangent vector $\displaystyle \ve{k}=\frac{\ve{R}}{R}$, where $\ve{R} = \ve{x}_{\obs} - \ve{x}_{\source}$ and the absolute 
value is $R = |\ve{R}|$. Furthermore, we define the unit tangent vector at remote past:  
$\displaystyle \ve{\sigma} = \lim_{t\rightarrow - \infty} \frac{\dot{\ve{x}}(t)}{c}$.

Up to post-Newtonian order, the differential equation (\ref{geodesic-post-Newtonian}) can be solved 
analytically. The solution for the transformation between $\ve{n}$ and $\ve{k}$ reads 
\begin{eqnarray}
\ve{n} &=& \ve{k} - (1 + \gamma) \,\frac{m}{d^{\;\prime}} \,\frac{\ve{d^{\;\prime}}}{d^{\;\prime}} \,
\frac{x_{\source}\,x_{\obs} - \ve{x}_{\source}\cdot \ve{x}_{\obs}}{R\,x_{\obs}} + {\cal O} \left(m^2\right),
\label{standard_25_A}
\end{eqnarray}

\noindent
in terms of the coordinate-independent impact vector $\ve{d}^{\;\prime}$, cf. Eq.~(57) of \cite{Article_Klioner_Zschocke}:
\begin{eqnarray}
\ve{d}^{\;\prime} &=&
\lim_{t\rightarrow - \infty} \ve{\sigma} \times \left(\ve{x} (t) \times \ve{\sigma}\right).
\label{impact_B}
\end{eqnarray}

\noindent
This impact parameter is identical to Chandrasekhar's impact parameter \cite{Article_Klioner_Zschocke,Report4}, 
that means in vectorial form $\ve{d}^{\;\prime} = \frac{\displaystyle \ve{L}}{\displaystyle E}$,
where $\ve{L}$ is the orbital three-momentum and $E$ is the energy of the photon on the light-trajectory; 
cf. Eq.~(215) in chapter 20 of \cite{Chandrasekhar1983}.

By means of $\sin \varphi = \left|\ve{n} \times \ve{k}\right|$, we find the light-deflection
angle $\varphi = \delta(\ve{n},\ve{k})$ in post-Newtonian approximation:
\begin{eqnarray}
\varphi &=& (1 + \gamma) \,\frac{m}{d^{\;\prime}}\,
\frac{x_{\source}\,x_{\obs} - \ve{x}_{\source}\cdot \ve{x}_{\obs}}{R\,x_{\obs}} + {\cal O} \left(m^2\right).
\label{standard_25_B}
\end{eqnarray}

\noindent
Note that $\displaystyle \frac{x_{\source}\,x_{\obs} - \ve{x}_{\source}\cdot \ve{x}_{\obs}}{R\,x_{\obs}}\le 2$, and 
therefore $\displaystyle \varphi \le \frac{4\,m}{d^{\;\prime}}$. 
One problem of post-Newtonian solution (\ref{standard_25_A}) or (\ref{standard_25_B}) is, that one can only state that the 
neglected terms are of order ${\cal O} \left(m^2\right)$, but their magnitude remains unclear. In order to make a 
statement about the upper magnitude of the higher order terms one needs to consider the geodesic equation in 
post-post-Newtonian approximation.

\section{Post-post Newtonian approximation\label{post_post_Newtonian}}

Now we will consider the trajectory of a light-signal in post-post-Newtonian Schwarzschild metric:
\begin{eqnarray} 
g_{00} &=& - 1 + 2\,a - 2\,\beta\,a^2 + {\cal O} \left(c^{-6}\right)\;,\quad g_{i0} = 0\,,
\nonumber\\
g_{ij} &=&\delta_{ij} + 2\,\gamma\,a\,\delta_{ij} + \epsilon \left(\delta_{ij} + \frac{x^i\,x^j}{x^2}\right) a^2
+ {\cal O} \left(c^{-6}\right). 
\label{ppN_metric_5}
\end{eqnarray}

\noindent
The geodesic equation of light-propagation in post-post-Newtonian approximation is given by \cite{Article_Klioner_Zschocke}
\begin{eqnarray}
\ddot{\ve{x}} &=&
-(1+\gamma)\,c^2\,\frac{a\,\ve{x}}{x^2}
+2\,(1+\gamma)\,\frac{a\,\dot{\ve{x}}\,(\dot{\ve{x}}\cdot\ve{x})}{x^2}
\nonumber
\\
&& \hspace{-0.5cm} +2\,c^2\,\left(\beta-\epsilon+2\,\gamma\,
(1+\gamma)\right)\,\frac{a^2\,\ve{x}}{x^2}
+2\,\epsilon\,\frac{a^2\,\ve{x}\,(\dot{\ve{x}}\cdot\ve{x})^2}{x^4}
\nonumber
\\
&& \hspace{-0.5cm} + 2\,(2(1-\beta)+\epsilon-2\,\gamma^2)\,
\frac{a^2\,\dot{\ve{x}}\,(\dot{\ve{x}}\cdot\ve{x})}{x^2}
+{\cal O} \left(c^{-4}\right).
\label{geodesic-post-post-Newtonian}
\end{eqnarray}

\noindent
The parameters $\beta$, $\gamma$ and $\epsilon$ characterize possible deviation of physical reality from general 
relativity theory (in general relativity $\beta = \gamma = \epsilon = 1$). The solution 
of (\ref{geodesic-post-post-Newtonian}) and the transformation between the unit vectors $\ve{n}$ and $\ve{k}$ in 
post-post-Newtonian order has been given in \cite{Article_Klioner_Zschocke}, cf. Eqs.~(108) and (109) ibid., and reads 
\begin{eqnarray}
\ve{n} &=& \ve{k}-(1+\gamma) \frac{m}{d^{\,\prime}} \frac{\ve{d}^{\,\prime}}{d^{\,\prime}} 
\frac{x_{\source}\,x_{\obs} - \ve{x}_{\source}\cdot\ve{x}_{\obs}}{R\,x_{\obs}}  
+ {\cal O} \left(\frac{m^2}{{d^{\,\prime}}^2}\right).
\label{lens_50}
\end{eqnarray}

\noindent
The terms of the order $\displaystyle {\cal O} \left(\frac{m^2}{{d^{\;\prime}}^2}\right)$ 
can be estimated to be smaller than or equal to $\displaystyle \frac{15\,\pi}{4}\,\frac{m^2}{{d^{\;\prime}}^2}$. 
From Eq.~(\ref{lens_50}) we obtain the expression 
\begin{eqnarray}
\varphi &=& 
(1 + \gamma)\,\frac{m}{d^{\;\prime}}\,\frac{x_{\source}\,x_{\obs} - \ve{x}_{\source} \cdot \ve{x}_{\obs}}{R\,x_{\obs}} 
+ {\cal O} \left(\frac{m^2}{{d^{\;\prime}}^2}\right).
\label{lens_155_A}
\end{eqnarray}

\noindent
The solutions (\ref{lens_50}) and (\ref{lens_155_A}) are identical to the post-Newtonian solution (\ref{standard_25_A}) 
and (\ref{standard_25_B}), respecticvely. This fact means that the post-post-Newtonian terms in the metric 
(\ref{ppN_metric_5}) and also the post-post-Newtonian terms in the geodesic equation (\ref{geodesic-post-post-Newtonian}) 
contribute only terms which can be estimated to be smaller than or equal to 
$\displaystyle \frac{15\,\pi}{4}\,\frac{m^2}{{d^{\;\prime}}^2}$. Therefore, the only difference between 
(\ref{lens_155_A}) and (\ref{standard_25_B}) here is, that the post-post-Newtonian approximation allows 
to make a statement about the upper magnitude of the regular post-post-Newtonian terms.

\section{Generalized lens equation\label{generalized_lens}}

Usually, in practical astrometry the position of observer $\ve{x}_{\obs}$ and the position of light-deflecting body 
is known (here, the center of massive body coincides with the coordinate center), 
but the impact parameter $d^{\;\prime}$ is not accessible. Therefore, the solutions (\ref{lens_50}) or 
(\ref{lens_155_A}) are not applicable in the form presented. Instead, one has to rewrite these solutions in terms of 
the impact vector of the unperturbed light ray  
\begin{eqnarray}
\ve{d} &=& \ve{k} \times \left( \ve{x}_{\obs} \times \ve{k} \right).
\label{impact_vector_5}
\end{eqnarray}

\noindent
For that one needs a relation between impact vector $\ve{d}^{\;\prime}$ defined in Eq.~(\ref{impact_B}) 
and impact vector $\ve{d}$ defined in Eq.~(\ref{impact_vector_5}). 
Such a relation has been given in \cite{Article_Klioner_Zschocke}, cf.~(62) ibid., and reads
(note, $d^{\;\prime} = d + {\cal O} \left(m\right)$):
\begin{eqnarray}
d^{\;\prime} &=& d + (1+\gamma)\,\frac{m}{d^{\prime}}\,\frac{x_{\source}+x_{\obs}}{R}\,
\frac{x_{\source}\,x_{\obs} - \ve{x}_{\source} \cdot\ve{x}_{\obs}}{R}
+ {\cal O} \left(m^2\right).
\label{coordinate_independent_15}
\end{eqnarray}

\noindent
Eq.~(\ref{coordinate_independent_15}) represents an quadratic equation for $d^{\;\prime}$, and these both 
solutions correspond to the two possible light-trajectories. 
A comparison of (\ref{coordinate_independent_15}) with (\ref{lens_155_A}) yields the relation 
\begin{eqnarray}
d^{\;\prime} &=& d + x_{\obs}\,\varphi + \frac{x_{\source} +
x_{\obs}-R}{R}\,x_{\obs}\,\varphi + {\cal O} \left(m^2\right)\,,
\label{simplest_form_5}
\end{eqnarray}

\noindent
where $\varphi$ is given by Eq.~(\ref{lens_155_A}) and we have separated a term 
$\displaystyle \frac{x_{\source} + x_{\obs} - R}{R}\,x_1\,\varphi$ which can be shown to contribute to the 
light-deflection $\varphi$ only to order $\displaystyle {\cal O} \left(\frac{m^2}{{d^{\;\prime}}^2}\right)$. 
By inserting (\ref{simplest_form_5}) into (\ref{lens_155_A}) we obtain an quadratic equation which has the solution  
\begin{eqnarray}
\varphi_{1,2} &=& \frac{1}{2} \left( \sqrt{\frac{d^2}{x_{\obs}^2}
+ 4\,(1+\gamma)\,\frac{m}{x_{\obs}}\,\frac{x_{\source}\,x_{\obs} - \ve{x}_{\source} \cdot\ve{x}_{\obs}}{R\;x_{\obs}}}
\mp \frac{d}{x_{\obs}} \right) + {\cal O} \left(\frac{m^2}{{d^{\;\prime}}^2}\right).
\label{simplest_form_25}
\end{eqnarray}

\noindent
The solution with the upper (lower) sign is denoted by $\varphi_1$ ($\varphi_2$). For astrometry the solution $\varphi_1$ 
can be considered to be the more relevant solution, because $\varphi_2$ represents the second image of 
one and the same source. 
One can show, that the terms $\displaystyle {\cal O} \left(\frac{m^2}{{d^{\;\prime}}^2}\right)$ 
are smaller than or equal to $\displaystyle \frac{15\,\pi}{4}\,\frac{m^2}{{d^{\;\prime}}^2}$. 
Equation (\ref{simplest_form_25}) represents the generalized lens equation. This equation is applicable not only
for any configurations where the post-Newtonian approach, the post-post-Newtonian approach, or
the classical lens equation is valid, but also for all those extreme configurations given
by the points (1) - (3) in the introductory section.
Especially, it allows an analytical investigation of light-deflection in binary systems \cite{Zschocke_Binaries}. 
In the following Section we will show that the formula (\ref{simplest_form_25}) represents a link between 
standard post-Newtonian approach and classical lens equation.

\section{Discussion of generalized lens equation\label{discussion}}

\subsection{Comparison with standard post-Newtonian and post-post-Newtonian approach}

In this Section we compare the generalized lens equation (\ref{simplest_form_25})
with the standard post-Newtonian and post-post-Newtonian approach of light-deflection.
A series expansion of the solution $\varphi_1$ in Eq.~(\ref{simplest_form_25}) for $d \gg m$ yields
\begin{eqnarray}
\varphi_1 &=& \varphi_{\rm pN} + \varphi_{\rm ppN} + {\cal O} \left(m^3\right) 
+ {\cal O} \left(\frac{m^2}{{d^{\;\prime}}^2}\right), 
\label{post_post-Newtonian_5}
\end{eqnarray}

\noindent
with
\begin{eqnarray}
\varphi_{\rm pN} &=& (1+\gamma)\,\frac{m}{d}\,
\frac{x_{\source}\,x_{\obs} - \ve{x}_{\source} \cdot\ve{x}_{\obs}}{R\,x_{\obs}}
\le 4\,\frac{m}{d}\,,
\label{post_post-Newtonian_10}
\\
\nonumber\\
\varphi_{\rm ppN} &=& - (1+\gamma)^2\,\frac{m^2}{d^2}\,
\frac{\left(x_{\source}\,x_{\obs} - \ve{x}_{\source} \cdot\ve{x}_{\obs}\right)^2}{R^2\;d\;x_{\obs}}
\le 16\,\frac{m^2}{d^2}\,\frac{x_{\obs}}{d}\,.
\label{post_post-Newtonian_15}
\end{eqnarray}

\noindent
Expression (\ref{post_post-Newtonian_10}) is called {\it standard post-Newtonian solution}, cf. Eq.~(24) in 
\cite{Article_Klioner_Zschocke}. The expression (\ref{post_post-Newtonian_15}) is just the 'enhanced' 
post-post-Newtonian term, cf. Eqs.~(3) and (4) in \cite{Report5}. 
The 'enhanced term' can be arbitrarily large for small $d$ and large $x_{\obs}$. 
That is the reason why the standard post-Newtonian and post-post-Newtonian solution is not applicable for extreme 
configurations like binary stars. The term ${\cal O} \left(m^3\right)$ will be discussed below, 
see Eq.~(\ref{post_post-post-Newtonian}); here it is only essential to realize that this term can be larger 
than the neglected terms $\displaystyle {\cal O}\left(\frac{m^2}{{d^{\;\prime}}^2}\right)$.  

\subsection{Comparison of generalized lens equation and classical lens equation \label{classical_lens_B}}

\begin{figure}[!h]
\caption{A geometrical representation of classical lens.}
\vspace{10pt}
\begin{indented}
\item[]
\includegraphics[scale=0.8]{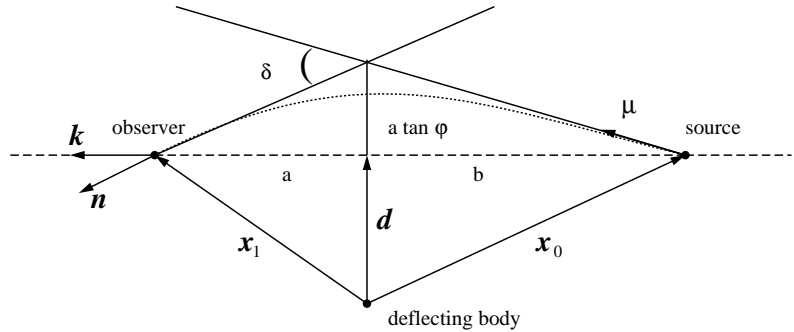}
\label{Fig: Lens}
\end{indented}
\end{figure}

The standard weak-field lens equation is usually called {\it classical lens equation} and given, for instance, in Eq.~(67) 
in \cite{exact_lens3} or Eq.~(24) in \cite{exact_lens4}. Let us briefly reconsider the classical lens equation. 
According to the scheme in Figure~\ref{Fig: Lens}, we obtain the following geometrical relations
\begin{eqnarray}
\varphi + \psi &=& \delta \,,
\label{lens_5}
\\
a\, \tan \varphi &=& b\, \tan \psi\,.
\label{lens_10}
\end{eqnarray}

\noindent
Here, the angles are $\psi = \delta (\ve{\mu}, \ve{k})$ and $\delta =
\delta (\ve{n}, \ve{\mu})$, where $\ve{\mu} = \frac{\dot{\ve{x}}
(t_0)}{|\dot{\ve{x}} (t_0)|}$ is the unit tangent vector at the position
of the source in the direction of the propagation of the light signal. 
If the source and observer are infinitely far from the massive body, then the total light-deflection angle
$\delta = \delta \left(\ve{n},\ve{\mu}\right)$ in Schwarzschild metric reads \cite{Bodenner_Will}
\begin{eqnarray}
\delta &=& 2\,\left(1 + \gamma\right) \frac{m}{d^{\;\prime}} + {\cal O} \left(\frac{m^2}{{d^{\;\prime}}^2}\right),
\label{lens_15}
\end{eqnarray}

\noindent
which is a coordinate independent result. The terms of order 
$\displaystyle {\cal O}\left(\frac{m^2}{{d^{\;\prime}}^2}\right)$ can be estimated to be smaller than or equal to 
$\displaystyle \frac{15\,\pi}{4}\,\frac{m^2}{{d^{\;\prime}}^2}\,$, see \cite{Bodenner_Will}. In classical lens approach, 
the approximation $d^{\;\prime} \simeq d + a\,\tan \varphi$ is used, see Figure~\ref{Fig: Lens}. 
Inserting this relation into (\ref{lens_15}), by means of geometrical relations (\ref{lens_5}) and (\ref{lens_10}), and 
using $\tan \varphi = \varphi + {\cal O} (\varphi^3)$ and $\tan \psi = \psi + {\cal O} (\varphi^3)$, we obtain the 
quadratic equation
\begin{eqnarray}
\varphi^2 + \frac{d}{a}\,\varphi - 2\,(1 + \gamma)\,\frac{m}{a}\,\frac{b}{a +b} &=& 0 \,.
\label{lens_20}
\end{eqnarray}

\noindent
The solution of Eq.~(\ref{lens_20}) is the classical lens equation:
\begin{eqnarray}
\varphi_{1,2}^{\rm class} &=& \frac{1}{2} \left( \sqrt{\frac{d^2}{a^2}
+ 8 (1 + \gamma)\,\frac{m}{a}\,\frac{b}{a + b}} \mp \frac{d}{a} \right),
\label{lens_25}
\end{eqnarray}

\noindent
which is valid in case of $a,b \gg d$; the solution with the upper (lower) sign is denoted by 
$\varphi_1^{\rm class}$ ($\varphi_2^{\rm class}$). 

It should be noticed that in (\ref{lens_25}) not only the light deflection angle $\varphi$ is
assumed to be small, but also the source and observer are assumed to be far from the massive body,
i.e. $\delta (\ve{x}_0, \ve{x}_1) \simeq \pi$; note that due to that fact equation (\ref{lens_25}) agrees with
the classical lens equation (67) in \cite{exact_lens3}. 
Therefore, the classical lens equation is not applicable for extreme configurations like binary systems 
or light deflection of moons at their giant planets of solar system.

It can easily be shown that the classical lens equation (\ref{lens_25}) follows straightforward 
from the generalized lens equation (\ref{simplest_form_25}). That means, if we rewrite (\ref{simplest_form_25}) in terms 
of $a = \ve{k}\cdot \ve{x}_{\obs}$ and $b = - \ve{k}\cdot \ve{x}_{\source}$ and perform a corresponding series expansion 
of generalized lens equation (\ref{simplest_form_25}) then we just obtain the classical lens equation (\ref{lens_25}) as 
the leading term in this series. 

Furthermore, in the limit $d \rightarrow 0$, known as Einstein ring solution, the generalized lens equation 
(\ref{simplest_form_25}) and the classical lens equation (\ref{lens_25}) yield the same result:  
\begin{eqnarray}
\lim_{d \rightarrow 0} \varphi_{1,2} &=& \lim_{d \rightarrow 0}\, \varphi_{1,2}^{\rm class} 
= \sqrt{2\left(1+\gamma\right)\,\frac{m}{x_{\obs}}\,
\frac{x_{\source}}{x_{\source}+x_{\obs}}}\,.
\label{d_0_10}
\end{eqnarray}

\noindent
Finally, we note that in the extreme configuration $b = 0$ (in this limit $\varphi_2$ does not exist) we obtain 
from (\ref{simplest_form_25}) the result 
\begin{eqnarray}
\lim_{b \rightarrow 0}\varphi_1 &=& \frac{1}{2}\left(\sqrt{\frac{d^2}{x_{\obs}^2}+ 4\left(1+\gamma\right)\frac{m}{x_{\obs}}\,
\frac{d\,a}{\left(x_{\obs}+d\right)\,x_{\obs}}}- \frac{d}{x_{\obs}}\right) 
 \le \sqrt{\left(1+\gamma\right)\frac{m}{x_{\obs}}}\,,
\label{limit_b_0}
\end{eqnarray}

\noindent
while the classical lens equation yields simply $\varphi_1^{\rm class}=0$. Obviously, in the limit $a\rightarrow 0$ 
the expression (\ref{limit_b_0}) yields zero as it has to be because in this 
limit the distance between source and observer vanishes, that means no 
light deflection.

\subsection{Comparison with exact solution}

The accuracy of (\ref{simplest_form_25}) and the stated estimate that the neglected terms are smaller than or 
equal to $\displaystyle \frac{15\,\pi}{4}\,\frac{m^2}{{d^{\;\prime}}^2} $ has also been confirmed by a comparison with 
the exact numerical solution of (\ref{exact_Schwarzschild_10}). 
 
\begin{figure}[!h]
\begin{center}
\includegraphics[scale=0.25,angle=270]{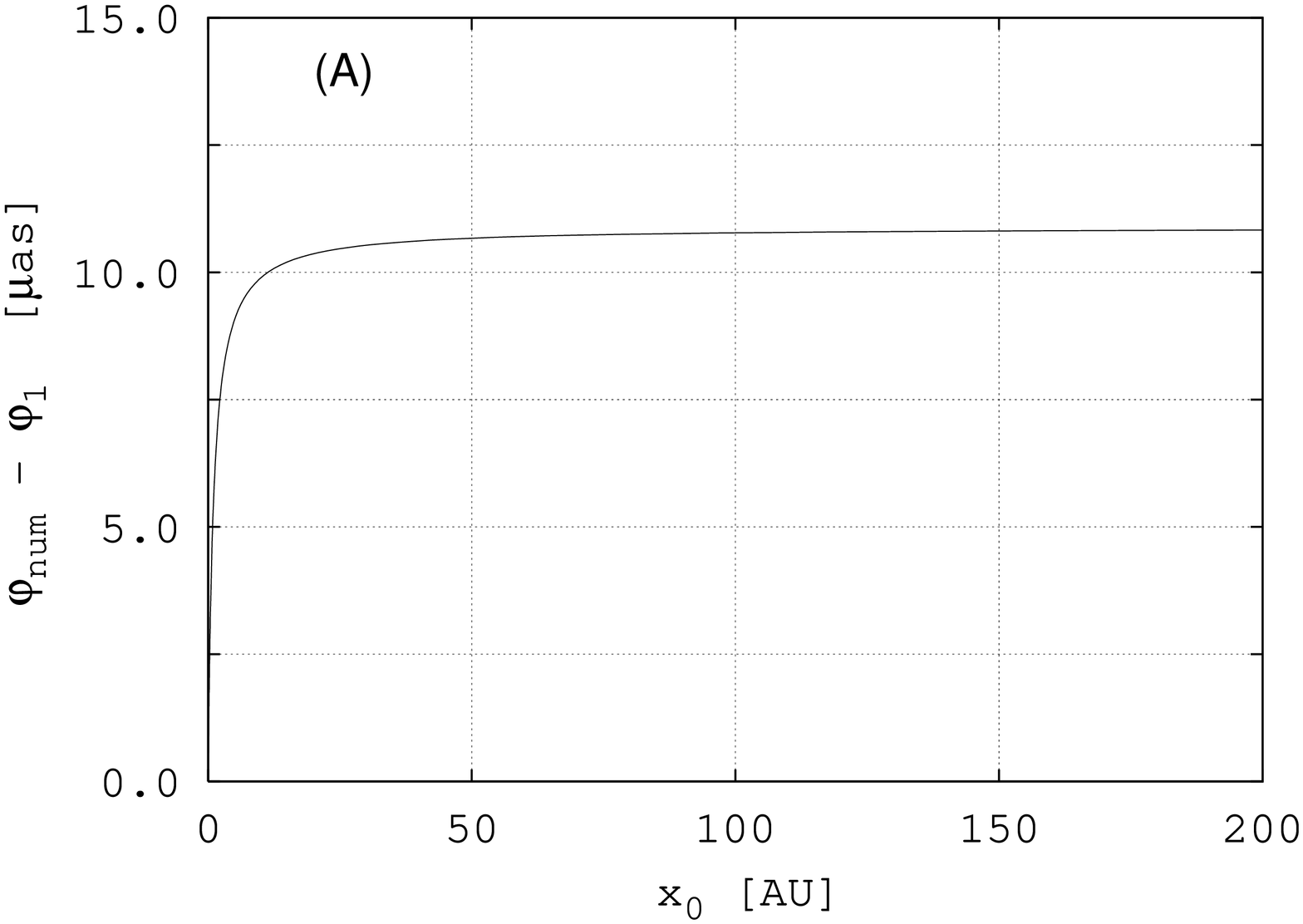}
\includegraphics[scale=0.25,angle=270]{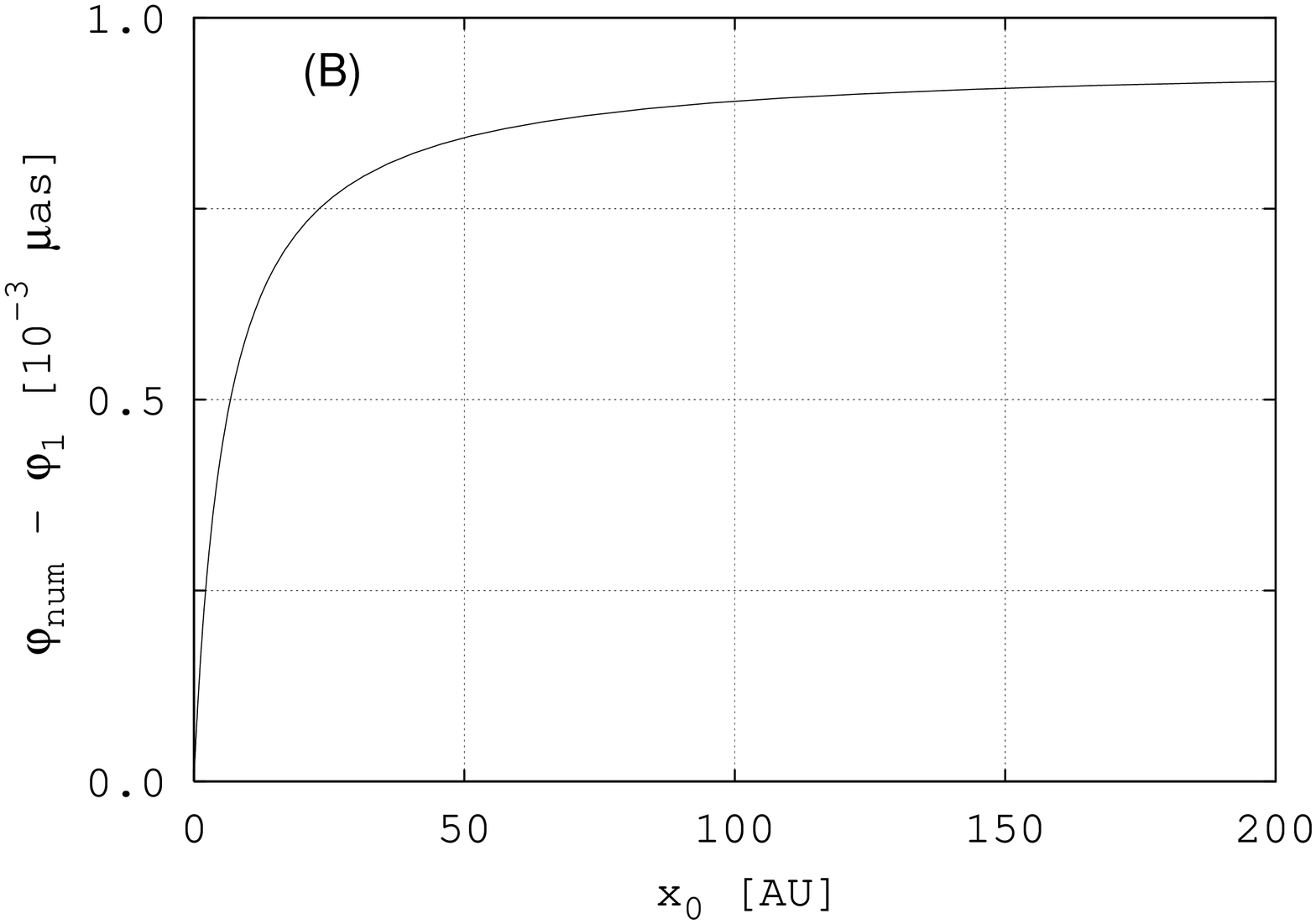}
\caption{Comparison of solution $\varphi_1$ of generalized lens equation (\ref{simplest_form_25}) 
with exact numerical solution $\varphi_{\rm num}$ for the case of a grazing ray at Sun (A) 
($\ve{x}_1 = (-1\,{\rm a.u.},0,0), m_{\odot}=1476.6\,{\rm m}$, $d^{\;\prime}=696.0 \times 10^6\,{\rm m}$) and Jupiter (B) 
($\ve{x}_1 = (- 6.0\,{\rm a.u.},0,0), m_{\jupiter}=1.40987\,{\rm m}$, $d^{\;\prime}=71.492 \times 10^6\,{\rm m}$), 
where ${\rm a.u.} = 1.496 \times 10^{11}\,{\rm m}$ denotes astronomical unit.}
\label{Fig: Comparison3}
\end{center}
\end{figure}

For that, we have solved the geodesic equation (\ref{exact_Schwarzschild_10}) in 
Schwarzschild metric by numerical integrator ODEX \cite{ODEX} for several extreme astrometrical configurations.
Using forth and back integration a numerical accuracy of at least $10^{-24}$ in the components of position and velocity
of the photon is guaranteed. Thus, the numerical integration can be considered as an exact solution of geodesic equation 
which is denoted by $\varphi_{\rm num}$. 
This numerical approach has been described in some detail in \cite{Article_Klioner_Zschocke}.
In all considered extreme configurations the validity of (\ref{simplest_form_25}) and the given 
estimate of neglected terms have been confirmed. As example, in Figure~\ref{Fig: Comparison3} 
we present the results for light-deflection of a grazing ray at Sun and Jupiter. 
These examples elucidate the fact that the accuracy of generalized lens equation (\ref{simplest_form_25}) is much beyond 
microarcsecond level of accuracy in case of light-deflection at giant planets. The reason for this fact is that 
$\displaystyle \frac{15\,\pi}{4}\,\frac{m^2}{{d^{\;\prime}}^2} \ll \mu{\rm as}$ for giant planets of the solar system; only 
in the vicinity of the Sun we have $\displaystyle \frac{15\,\pi}{4}\,\frac{m^2}{{d^{\;\prime}}^2} \sim 11\,\mu{\rm as}$. 

The accuracy shown in Figure~\ref{Fig: Comparison3} (B) is considerably better 
than the post-post-Newtonian solution investigated in detail in \cite{Article_Klioner_Zschocke,Report2}, 
cf. Figure~\ref{Fig: Comparison3} (B) with FIG.~$2$ in \cite{Report2}.
In order to understand the numerical difference between Figure~\ref{Fig: Comparison3} (B) and 
FIG.~$2$ in \cite{Report2}, we perform a further series expansion of Eq.~(\ref{simplest_form_25}) up to terms of 
order $m^4$, that means 
\begin{eqnarray}
\varphi_1 &=& \varphi_{\rm pN} + \varphi_{\rm ppN} + \varphi_{\rm pppN} 
+ {\cal O} \left(m^4\right) + {\cal O} \left(\frac{m^2}{{d^{\,\prime}}^2}\right),
\label{expansion_2}
\end{eqnarray}

\noindent
where the 'enhanced' terms beyond post-post-Newtonian terms are:
\begin{eqnarray}
\varphi_{\rm pppN} &=&  2\,\left(1+\gamma\right)^3\,\frac{m^3}{d^3}\,
\frac{\left(x_{\source}\,x_{\obs} - \ve{x}_{\source} \cdot\ve{x}_{\obs}\right)^3}{R^3\;d^2\;x_{\obs}}
\le 128\,\frac{m^3}{d^3}\,\frac{x_{\obs}^2}{d^2}\,.
\label{post_post-post-Newtonian}
\end{eqnarray}

\noindent
The given estimation in (\ref{post_post-post-Newtonian}) shows, that for large $x_1$ this term can be considerably larger 
than the neglected terms of order $\displaystyle {\cal O} \left(\frac{m^2}{{d^{\;\prime}}^2}\right)$. 
Moreover, the numerical difference between Figure~\ref{Fig: Comparison3} and FIG.~$2$ in \cite{Report2} 
is just given by the term (\ref{post_post-post-Newtonian}).

\section{Summary\label{summary}}

Modern astrometry has achieved a microarcsecond level of accuracy, e.g. astrometric missions SIM (NASA) or GAIA (ESA). 
A time-efficient data reduction implies the need of approximative and highly precise solutions for the light deflection 
on this level of accuracy. In our investigation we have suggested a generalized lens equation (\ref{simplest_form_25}) for weak gravitational fields of 
Schwarzschild metric and valid for finite distances of source and observer 
from the light deflecting body.
The derivation is based on the solution of geodesic equation (\ref{lens_155_A}) in 
post-Newtonian metric and Chandrasekhar's coordinate independent impact parameter $d^{\;\prime}$ (\ref{impact_B}) and 
its relation to the light-deflection angle $\varphi$ given in (\ref{simplest_form_5}). 
The neglected terms in (\ref{simplest_form_25}) can be estimated to be smaller than or 
equal to $\displaystyle \frac{15\,\pi}{4}\,\frac{m^2}{{d^{\;\prime}}^2}$. The accuracy of generalized lens equation 
(\ref{simplest_form_25}) is considerably better than the standard post-Newtonian and post-post-Newtonian approach, 
which has been investigated in some detail in \cite{Article_Klioner_Zschocke,Report2} and the reason for this fact 
has been pointed out. 

The generalized lens equation (\ref{simplest_form_25}) satisfies three conditions formulated in the introductory Section. 
Moreover, we have shown that in the appropriate limits we obtain the post-Newtonian terms, 'enhanced' 
post-post-Newtonian terms and the classical lens equation. Thus, the generalized lens equation 
(\ref{simplest_form_25}) provides also a link between these essential approaches to determine the light-deflection. 
Numerical investigations have confirmed the analytical results obtained. 

The generalized lens equation (\ref{simplest_form_25}) allows an analytical understanding and investigation 
of light-deflection in extreme astrometric configurations. Especially, the determination of light-deflection in 
binary systems using of generalized lens equation (\ref{simplest_form_25}) 
has been investigated in \cite{Zschocke_Binaries}.

\section*{Acknowledgements}

This work was partially supported by the BMWi grants 50\,QG\,0601 and
50\,QG\,0901 awarded by the Deutsche Zentrum f\"ur Luft- und Raumfahrt
e.V. (DLR). Enlighting discussions with Prof. Sergei A. Klioner, 
Prof. Michael Soffel and Prof. Chongming Xu are greatfully acknowledged.

\end{document}